\documentclass[twocolumn,astrosymb,tighten,twocolappendix]{aastex631}

\usepackage{threeparttable}
\usepackage{amsmath}
\usepackage{url}

\newcommand{\HII}{\mbox{\ion{H}{2}}} 

\newcommand{\diff}{\mathrm{d}}

\graphicspath{{./}{figures/}}

\begin{document}
\begin{sloppypar}

\title{Deciphering Radio Emissions from Accretion Disk Winds in Radio-Quiet Active Galactic Nuclei}

\author{Tomoya Yamada}
\affiliation{Department of Earth and Space Science, Graduate School of Science, Osaka University, 1-1, Machikaneyama, Toyonaka, Osaka560-0043, Japan}

\author{Nobuyuki Sakai}
\affiliation{Department of Earth and Space Science, Graduate School of Science, Osaka University, 1-1, Machikaneyama, Toyonaka, Osaka560-0043, Japan}

\author[0000-0002-7272-1136]{Yoshiyuki Inoue}
\affiliation{Department of Earth and Space Science, Graduate School of Science, Osaka University, 1-1, Machikaneyama, Toyonaka, Osaka560-0043, Japan}
\affiliation{Interdisciplinary Theoretical \& Mathematical Science Program (iTHEMS), RIKEN, 2-1 Hirosawa, 351-0198, Japan}
\affiliation{Kavli Institute for the Physics and Mathematics of the Universe (WPI), UTIAS, The University of Tokyo, 5-1-5 Kashiwanoha, Kashiwa, Chiba 277-8583, Japan}

\author[0000-0003-2475-7983]{Tomonari Michiyama}
\affiliation{Faculty of Information Science, Shunan University, 843-4-2 Gakuendai, Shunanshi, Yamaguchi 745-8566, Japan}

\begin{abstract}
Unraveling the origins of radio emissions from radio-quiet active galactic nuclei (RQ AGNs) remains a pivotal challenge in astrophysics. One potential source of this radiation is the shock interaction between AGN disk winds and the interstellar medium (ISM). To understand this phenomenon,  we construct a spherical, one-zone, and self-similar expansion model of shock structure between ultra-fast outflows (UFOs) and the ISM. We then calculate the energy density distribution of non-thermal electrons by solving the transport equation, considering diffusive shock acceleration as the acceleration mechanism and synchrotron and inverse Compton cooling as the cooling mechanisms. Based on the derived energy distribution of non-thermal electrons, we model the radio synchrotron spectrum of shocked ISM. For the 15 nearby RQ AGNs hosting UFOs, we investigate shocked ISM parameters required to model their observed radio spectra, based on X-ray observations and measured UFO velocities. Radio spectra of 11 out of 15 nearby RQ AGNs would be explained by the AGN disk wind model. This is a compelling indication that shock interactions between AGN disk winds and the ISM could indeed be the source of their radio emissions. The typical predicted source size and magnetic field strength are several $100$ pc and $0.1$ mG, respectively. 
We also discuss whether our prediction can be tested by future radio observations.
\end{abstract}

\section{introduction}

Active galactic nuclei (AGNs) outshine their host galaxies, primarily due to the release of gravitational energy as matter accretes onto their central supermassive black holes (SMBHs). Based on their radio loudness, which is the ratio of radio to optical luminosity, AGNs are bifurcated into two primary categories: radio-loud (RL) and radio-quiet (RQ) \citep{Kellermann1989,Sikora2007}. The radio brightness of RL AGNs is attributed to their powerful relativistic jets \citep{Blandford2019}. In contrast, the source of radio emissions in RQ AGNs remains enigmatic \citep{Panessa2019}, despite them constituting approximately 90\% of the entire AGN population.

Several hypotheses have been proposed to elucidate the origins of radio emissions in RQ AGNs. These include star formation activities \citep{Sopp1991,Sargent2010}, diminutive jets \citep{Ho2001,Nagar2001}, AGN coronae \citep{Laor2008,Inoue2014,Raginski2016MNRAS.459.2082R, Doi2016,Inoue2018a}, and, AGN disk winds \citep{Wang2008,Jiang2010,Zakamska2014, Nims2015}. Of these, AGN disk winds, emanating from the AGN accretion disk, are particularly intriguing. Their kinetic energy can approach nearly 5\% of their bolometric luminosities \citep{Tombesi2012,Gofford2015}. It is widely believed that these winds can potentially suppress the star-forming activities of their host galaxies \citep{Dimatteo2005,Hopkins2009}.  Therefore, a comprehensive understanding of the interaction of AGN disk winds with the interstellar medium (ISM)  is imperative for a holistic understanding of galaxy formation.

Within the realm of AGN disk winds, ultra-fast outflows (UFOs) are characterized as the fastest and most powerful winds whose velocities reach $\sim10\%$ of the speed of light \citep{Tombesi2010,Tombesi2011,Tombesi2012}. UFOs are observed in around 40\% of nearby AGNs. They originate from regions ranging from $10^{1}~R_\mathrm{{g}}$ to $10^{4}~R_\mathrm{{g}}$$\sim10^{-2}~$pc away from the central SMBHs, where $R_{\mathrm{g}}$ is the gravitational radius \citep{Gofford2015,Mizumoto2018,Mizumoto2019}. Observational data suggests that the expansive molecular outflows can be attributed to the energy-conserving nature of UFOs \citep{Tombesi2015,Mizumoto2019a}. This suggests that, despite the significant spatial discrepancy between UFO launch sites and their host galaxies, UFOs would influence the AGN feedback process.

\citet{Zakamska2014} postulated a correlation between the ionized gas outflow velocities, inferred from blueshifted [O III] lines, and radio luminosity in RQ AGNs, attributing it to AGN wind shock interactions \citep[see also][]{Liao2024MNRAS.528.3696L}. Numerous theoretical explorations have delved into the connection between AGN disk wind activities and radio emissions  \citep{Wang2008,Jiang2010,Faucher-Giguere2012,Nims2015, Wang2015}. 
In these models, non-thermal particles emerge in the shock region due to diffusive shock acceleration, leading to synchrotron radiation in the radio spectrum. Yet, a comparison between the predicted radio spectrum from AGN wind shock and actual radio spectral data remains elusive.

In this study, we compute radio synchrotron spectra of AGN wind shock, incorporating particle acceleration and cooling dynamics at the shock interface between UFOs and the ISM.  We further scrutinize the AGN wind shock parameters necessary to mirror the radio observational data of nearby RQ AGNs exhibiting UFOs. Our model is described in Section \ref{sec:model}, followed by results in Section \ref{sec:result}. Subsequent discussions and conclusions are presented in Section \ref{sec:discussion} and \ref{sec: conclusion}, respectively.

\section{Model Description}
\label{sec:model}

\subsection{Dynamics of AGN Wind}
\label{subsec: kinetic model of expansion of shock}
We construct a dynamical model for spherically symmetric expanding UFOs, following \citet{Nims2015}. We posit that UFOs are continuously launched, subsequently colliding with the ISM, leading to an expanding shock front governed by a self-similar solution.

To derive the kinetic luminosity of UFOs, denoted as $L_{\mathrm{kin}}$, we suppose that the ratio $\tau_{\mathrm{UFO}}$ of the momentum radiated by AGNs is transferred to UFOs. Given the AGN luminosity $L_{\mathrm{AGN}}$, $L_{\mathrm{kin}}$ can be expressed as,
\begin{equation}
    \begin{split}
        L_{\mathrm{kin}} &= 5\times10^{44}~\mathrm{erg\,s^{-1}} \\
        &\quad\times \left(\frac{v_{\mathrm{UFO}}}{0.1~c}\right)
        \left(\frac{L_{\mathrm{AGN}}}{10^{46}~\mathrm{erg\,s^{-1}}}\right) \tau_{\mathrm{UFO}},
    \end{split}
    \label{eq:kinetic luminosity of UFO}
\end{equation}
where $c$ is the speed of light. $\tau_\mathrm{UFO}$ is still observationally uncertain (see e.g., \citealp{Mizumoto2019ApJ...871..156M}). It becomes $\sim20$ at molecular outflow scales with an outflow velocity of $\sim1000~\mathrm{km\,s^{-1}}$ \citep{Cicone2014}. For the sake of simplicity, this paper adopts $\tau_{\mathrm{UFO}}=1$ below. 

We set the ISM density $n_{\mathrm{H}}(R)$ as the following power-law distribution,
\begin{equation}
    n_{\mathrm{H}}(R) = n_{\mathrm{H},0} \left(
    \frac{R}{R_0}
    \right)^{-\alpha},
    \label{eq:ISM distribution}
\end{equation}
where $R$ is the distance from the central SMBH and $\alpha$ is the power law index. To model observations of molecular outflows whose velocity is $\sim1000~$km/s at $\sim1~$kpc, we take $n_{\mathrm{H},0}=10\,\mathrm{cm^{-3}}$ at $R_0=100 \,\mathrm{pc}$ and $\alpha=1$ \citep{Faucher-Giguere2012, Nims2015}.

We suppose that half of the injected kinetic energy goes to the kinetic energy of the expanding shock, and the rest goes to thermal energy. By taking into account the Coulomb cooling, the outflow is energy-conserving for the range of parameters we have used in this paper \citep{Faucher-Giguere2012}. The energy conservation equation after time $t$ from the initial launch time of UFOs can be written as
\begin{equation}
    \label{eq: energy conserving equation about shock}
    \frac{1}{2} L_{\mathrm{kin}} t = \frac{1}{2} M_{\mathrm{sh}} v_{\mathrm{sh}}^2.
\end{equation}
Here, $M_{\mathrm{sh}}$ is the swept-up ISM mass given as
\begin{align}
    \label{eq: mass of swept up matters}
    \begin{split}
        M_{\mathrm{sh}} &= 4\pi m_{\mathrm{p}}\int_0^{R_{\mathrm{sh}}}n_{\mathrm{H}}(R) R^2\diff R \\
        &\simeq 1.5 \times10^6 \,\mathrm{M_{\odot}}\, \left(\frac{n_{\mathrm{H},\,0}}{10\,\mathrm{cm^{-3}}}\right) \left( \frac{R_0}{100\,\mathrm{pc}}\right) \left(\frac{R_{\mathrm{sh}}}{100\,\mathrm{pc}}\right)^2,
    \end{split}
\end{align}
where $R_{\mathrm{sh}}$ and $m_{\mathrm{p}}$ are the shock front radius and the proton mass, respectively. The contribution of the mass of the UFO component is negligible because the mass outflow rate of UFOs ranges from $0.01$ to $1~\mathrm{M_{\odot}\,yr^{-1}}$ \citep{Gofford2015}.

From the self-similar solution, the velocity of the shock front $v_{\mathrm{sh}}$ is derived as follows
\begin{equation}
    \begin{split}
        &v_{\mathrm{sh}} = \frac{1}{2}\left\{6 \frac{(3-\alpha)L_{\mathrm{kin}}}{(5-\alpha)\pi R_0^{\alpha}n_{\mathrm{H},0}m_{\mathrm{p}}}\right\}^{1/3}R_{\mathrm{sh}}^{(\alpha-2)/3} \\
        &\approx 3400\,\mathrm{km\,s^{-1}}\times
        \left( \frac{6-2\alpha}{5-\alpha}\right)^{1/3} 
        \left(\frac{v_{\mathrm{UFO}}}{0.1~c}\right)^{1/3}
        \left(\frac{L_{\mathrm{AGN}}}{10^{46}~\mathrm{erg\,s^{-1}}}\right)^{1/3} \\
        &\left(\frac{R_0}{100\,\mathrm{pc}}\right)^{-\alpha/3}
        \left(\frac{n_{\mathrm{H},0}}{10\,\mathrm{cm^{-3}}}\right)^{-1/3}
        \left( \frac{R_{\mathrm{sh}}}{100\,\mathrm{pc}}\right)^{(\alpha-2)/3}.
    \end{split}
    \label{eq:shell velocity}
\end{equation}
This velocity range corresponds to the observation results of molecular outflow \citep{Feruglio2010,Cicone2014}. The Mach number of this forward shock can be calculated as follows
\begin{align}
    \mathcal{M} &= v_{\mathrm{sh}}\sqrt{\frac{3}{5} \frac{\mu m_{\mathrm{p}}}{k_{\mathrm{B}}T_{\mathrm{ISM}}}} \\
    \label{eq:mach number}
    &\sim 200\times \left(\frac{T_{\mathrm{ISM}}}{10^4~\mathrm{K}}
    \right)^{-1/2} \left( \frac{v_{\mathrm{sh}}}{3400~\mathrm{km/s}}
    \right),
\end{align}
where $\mu$, $k_{\mathrm{B}}$ and $T_{\mathrm{ISM}}$ are mean molecular weight, Boltzmann constant, and the temperature of ISM.

The dynamical timescale of the shock front is given by
\begin{equation}
    \begin{split}
        &t_{\mathrm{dyn}} =\frac{R_{\mathrm{sh}}}{v_{\mathrm{sh}}} \\
        &\simeq 3 \times 10^4\, \mathrm{yr}
        \left( \frac{6-2\alpha}{5-\alpha}\right)^{-1/3} 
        \left(\frac{v_{\mathrm{UFO}}}{0.1~c}\right)^{-1/3}
        \left(\frac{L_{\mathrm{AGN}}}{10^{46}~\mathrm{erg\,s^{-1}}}\right)^{-1/3} \\
        &\left(\frac{R_0}{100\,\mathrm{pc}}\right)^{\alpha/3}
        \left(\frac{n_{\mathrm{H},0}}{10\,\mathrm{cm^{-3}}}\right)^{1/3}
        \left( \frac{R_{\mathrm{sh}}}{100\,\mathrm{pc}}\right)^{(5-\alpha)/3}.
    \end{split}
    \label{eq:dynamical timescale}
\end{equation}
This dynamical timescale aligns with observations suggesting that the active phase of AGNs spans approximately $10^5$~yr \citep{Schawinski2015}.

\subsection{Acceleration and Cooling of Non-thermal Electrons}
\label{subsec:acceleration and cooling of non-thermal electrons}
We suppose that electrons in the shocked ambient medium (SAM) undergo acceleration through the diffusive shock acceleration (DSA) process at shock front region \citep{Krymskii1977,Axford1977,Bell1978,Bell1978a,Blandford1978}. Recent studies, such as by \citet{Peretti2023}, have also suggested the shocked wind (SW) as a potential site for particle acceleration. In this study, our focus remains on the shock front region, especially given the high Mach number achieved there, as indicated in Eq.~\ref{eq:mach number}.

We define the energy density spectrum for the injection of non-thermal electrons, $q(\mathrm{\gamma_{\mathrm{e}}})$, as
\begin{equation}
    q(\gamma_{\mathrm{e}}) = q_0\gamma_{\mathrm{e}}^{-p} \exp{\left(-\frac{\gamma_{\mathrm{e}}}{\gamma_{\mathrm{cut}}}\right)}.
    \label{eq:injection function}
\end{equation}
Here, $q_0$, $\gamma_{\mathrm{e}}$, and $\gamma_{\mathrm{cut}}$ represent the normalization constant, the electron Lorentz factor, and the cut-off electron Lorentz factor, respectively. The normalization constant $q_0$ is determined by assuming that a fraction $\xi_{\mathrm{e}}$ of the UFO kinetic energy is dissipated into non-thermal electrons as
\begin{equation}
    \xi_{\mathrm{e}}L_{\mathrm{kin}} = \frac{4}{3}\pi R_{\mathrm{sh}}^3m_{\mathrm{e}}c^2 \int q(\gamma_{\mathrm{e}}) \gamma_{\mathrm{e}} \diff \gamma_{\mathrm{e}},
    \label{eq:normalization}
\end{equation}
where $m_{\mathrm{e}}$ is the rest mass of electrons. We set $\xi_{\mathrm{e}}$ to $0.01$, based on particle-in-cell simulations \citep{Park2015PhRvL.114h5003P} and observations of supernova remnants  \citep{Ackermann2013Sci...339..807A}.

The cut-off Lorentz factor $\gamma_{\mathrm{cut}}$ is determined by the balance between the acceleration time $t_{\mathrm{acc}}$ and the cooling time $t_{\mathrm{cool}}$. Assuming a Bohm-like diffusion, DSA acceleration timescale is given by \citep{OCDrury1983},
\begin{align}
    \label{eq:acceleration timescale}
    \begin{split}
        t_{\mathrm{acc}} &= \frac{8}{3}
        \frac{\eta m_{\mathrm{e}} c^3}
        {e B v_{\mathrm{sh}}^2}\gamma_{\mathrm{e}} \\
        &\simeq 1~\mathrm{yr} \left(\frac{\gamma_{\mathrm{e}}}{10^6}
        \right)\left(
        \frac{B}{0.1~\mathrm{mG}}\right)^{-1} \left(
        \frac{v_{\mathrm{sh}}}{3400~\mathrm{km\,s^{-1}}}\right)^{-2},
    \end{split}
\end{align}
where $\eta$, $e$, and $B$ are the gyrofactor, the charge of electrons, and the magnetic field strength respectively. In this paper, $\eta$ is fixed to unity, which does not significantly change our radio spectrum results. This is because the electrons, emitting synchrotron photons in our region of interest (the cm-wave region) have energies much lower than the cutoff energy.

For cooling timescales, we consider both synchrotron and inverse Compton (IC) cooling mechanisms. The contribution of free-free emission is negligible, as discussed in \citep{Nims2015}. The synchrotron cooling timescale is
\begin{align}
    \label{eq:synchrotron timescale}
        t_{\mathrm{syn}} &= \frac{3m_{\mathrm{e}}c^2}{4c\sigma_{\mathrm{T}}}\gamma_{\mathrm{e}}^{-1} U_{B}^{-1} \\
        &\simeq 3\times10^{2} \mathrm{yr}\left(\frac{\gamma_{\mathrm{e}}}{10^6}\right)^{-1} \left(
        \frac{B}{0.1\,\mathrm{mG}}\right)^{-2},
\end{align}
where $\sigma_{\mathrm{T}}$ and $U_{B}$ are the Thomson scattering cross section and the magnetic field energy density, respectively \citep{Rybicki1986}. 

The IC cooling timescale is 
\begin{equation}
    \label{eq:inverse Compton timescale}
    t_{\mathrm{IC}} = \frac{3m_{\mathrm{e}}c^2}{4c\sigma_{\mathrm{T}}}\gamma_{\mathrm{e}}^{-1} U_{\mathrm{ph}} ^{-1} F_{\mathrm{KN}}(\gamma_{\mathrm{e}})^{-1},
\end{equation}
where $U_{\mathrm{ph}}$ is the photon field energy density \citep{Rybicki1986}. $F_{\mathrm{KN}}(\gamma_{\mathrm{e}})$ represents the Klein-Nishina (KN) effect as 
\begin{equation}
    F_{\mathrm{KN}}(\gamma_{\mathrm{e}}) = \frac{1}{U_{\mathrm{ph}}} \int_{\epsilon_{\mathrm{0,min}}} ^{\epsilon_{\mathrm{0,max}}}
    f_{\mathrm{KN}}(\Tilde{b}) u(\epsilon_{0}) \diff \epsilon_0,
    \label{eq:Klein-Nishina function}
\end{equation}
where $\Tilde{b}$, $u(\epsilon)$, and $\epsilon_{0}$ are $\Tilde{b} = 4\gamma_{\mathrm{e}} \epsilon_{0}$, the energy density of soft photon per unit energy, and the energy of the soft photon normalized by the rest energy of electrons, respectively \citep{Moderski2005a}.
For simplicity, we assume the soft photon field is isotropic black body radiation emanating from AGN accretion disks with a temperature $T_{\mathrm{disk}}$ of 10~eV. The function $f_{\mathrm{KN}}$ is detailed in Sec.~2.1 in \citet{Moderski2005a}. 

In the Thomson scattering regime, where $f_{\mathrm{KN}}=1$, we have
\begin{align}
    t_{\mathrm{IC}} &= \frac{3m_{\mathrm{e}}c^2}{4c\sigma_{\mathrm{T}}}\gamma_{\mathrm{e}}^{-1} U_{\mathrm{ph}} ^{-1} \\
    &\simeq 4 \times 10^4~\mathrm{yr} \left(
    \frac{R_{\mathrm{sh}}}{100~\mathrm{pc}}\right)^2 \left(
    \frac{L_{\mathrm{AGN}}}{10^{46}~\mathrm{erg\,s^{-1}}}\right)^{-1} \left(
    \frac{\gamma_{\mathrm{e}}}{10^{2}}\right)^{-1}.
\end{align}

\begin{figure}
    \centering
    \includegraphics[width=1\linewidth]{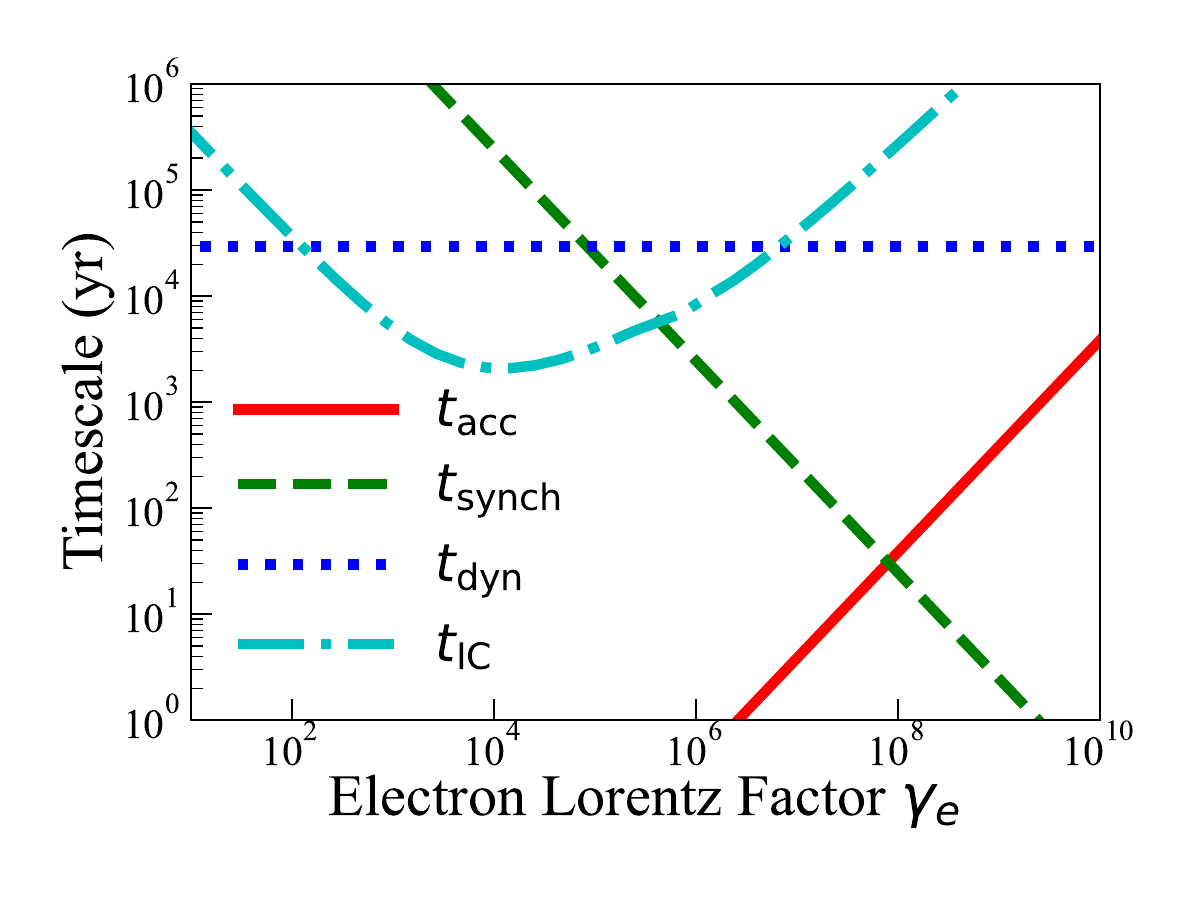}
    \caption{Acceleration (red solid), synchrotron cooling (green dashed), IC cooling (cyan dot-dashed), and dynamical (blue dotted) timescales versus electron Lorentz factor $\gamma_{\mathrm{e}}$. We set the bolometric luminosity of AGN as $L_{\mathrm{AGN}} = 10^{46} \mathrm{erg \, s^{-1}}$, the velocity of UFOs as $v_{\mathrm{UFO}} = 0.1 c$, the density of ISM at $100$ pc as $n_{\mathrm{H},\,0}=10\,\mathrm{cm^{-3}}$, the power law index of the spatial density distribution of the ISM as $\alpha=1$, size of the shock front as $R_{\mathrm{sh}}=100$ pc, magnetic field strength as $B=0.1$ mG, the gyrofactor as $\eta=1$, and the temperature of AGN accretion disk as $10$ eV. }
    \label{fig:cooling timescale}
\end{figure}

Figure~\ref{fig:cooling timescale} shows the acceleration and cooling timescales of electrons. The parameters used in these calculations are shown in Table~\ref{tab:fiducial parameter}. In the Thomson scattering regime, as shown in the figure, the IC radiation is the dominant cooling process, while, at higher energies, the synchrotron cooling becomes dominant due to the KN effect. Therefore, the cutoff energy is determined by the balance between $t_{\mathrm{acc}}$ and $t_{\mathrm{syn}}$. 

\begin{table*}
    \centering
    \caption{Fiducial value of the model parameters.}
    \begin{tabular}{l|ccc}
        \hline \hline 
        quantity & parameter & value & unit \\
        \hline
        AGN bolometric luminosity & $L_{\mathrm{AGN}}$ & $10^{46}$ & $\mathrm{erg\,s^{-1}}$  \\
        momentum ratio of UFO & $\tau_{\mathrm{UFO}}$ & 1 & --- \\
        velocity of UFOs & $v_{\mathrm{UFO}}$ & 0.1 & $c$   \\
        size of radio source & $R_{\mathrm{sh}}$ & 100 & pc  \\
        ISM density at distance of $100$ pc & $n_{H,0}$ & 10 & $\mathrm{cm^{-3}}$ \\
        power-law index of ISM spatial distribution & $\alpha$ & 1 & --- \\
        temperature of AGN accretion disk & $T_{\mathrm{disk}}$ & 10 & eV \\
        magnetic field strength & $B$ & $0.1$ & mG  \\
        power law index of injected non-thermal electrons & $p$ & $2$ & ---  \\
        energy fraction of non-thermal particles to UFOs & $\xi_{\mathrm{e}}$ & 0.01 & --- \\
        gyrofactor & $\eta$ & 1 & ---  \\
        size of \HII{} region & $D_{\mathrm{HII}}$ & $R_{\mathrm{sh}}$ & ---\\
        \hline
    \end{tabular}
    \label{tab:fiducial parameter}
\end{table*}

To derive the steady-state energy distribution of electrons $\diff n_{\mathrm{e}}/\diff \gamma_{\mathrm{e}}$, we consider the following transport equation:
\begin{align}
    \frac{\diff}{\diff \gamma_{\mathrm{e}}}\left(\dot{\gamma}_{\mathrm{cool}} \frac{\diff n_{\mathrm{e}}}{\diff \gamma_{\mathrm{e}}}\right)
    + \frac{1}{t_{\mathrm{dyn}}} \frac{\diff n_{\mathrm{e}}}{\diff \gamma_{\mathrm{e}}}
    &= q(\gamma_{\mathrm{e}}),
    \label{eq:Focker-Planck equation}
\end{align}
where $\dot{\gamma}_{\mathrm{cool}} = \dot{\gamma}_{\mathrm{syn}} + \dot{\gamma}_{\mathrm{IC}}$ is the total cooling rate of electrons, 
$\dot{\gamma}_{\mathrm{syn}} = \gamma_{\mathrm{e}}/t_{\mathrm{syn}}$ is the synchrotron cooling rate, and $\dot{\gamma}_{\mathrm{IC}} = \gamma_{\mathrm{e}}/t_{\mathrm{IC}}$ is the IC cooling rate, respectively. The analytical solution of this equation is given by \citep{Ginzburg1966}
\begin{align}
    \label{eq:energy distribution of non-thermal electron}
    \frac{\diff n_{\mathrm{e}}}{\diff \gamma_{\mathrm{e}}} &=
    \frac{1}{\dot{\gamma}_{\mathrm{cool}}}\int_{\gamma_{\mathrm{e}}}^{\infty}
    q(\gamma_{\mathrm{e}}')e^{-T(\gamma_{\mathrm{e}},\gamma_{\mathrm{e}}')} \diff \gamma_{\mathrm{e}}'\\
    \label{eq:function T in electron energy spectrum}
    T(\gamma_{1},\gamma_{2}) &= 
    \frac{1}{t_{\mathrm{dyn}}}\int_{\gamma_{1}}^{\gamma_{2}}
    \frac{1}{\dot{\gamma}_{\mathrm{cool}}}\diff \gamma_{\mathrm{e}}.
\end{align}
By employing Equation (\ref{eq:energy distribution of non-thermal electron}), we derive the electron spectrum, subsequently computing the synchrotron spectra.

\subsection{Synchrotron Radiation}
\label{subsec:Synchrotron radiation process}
We compute the synchrotron radiation power, $P(\nu, \, \gamma_{\mathrm{e}})$, under the tangled magnetic field configuration, following \citet{Aharonian2010a, Rybicki1986}.
\par
Synchrotron emissivity $j_{\nu}^{\mathrm{syn}}$ and the synchrotron absorption coefficient $\alpha_{\nu}^{\mathrm{syn}}$ are \citep{Rybicki1986}
\begin{align}
    \label{eq:sychrotron emissivity}
    j_{\nu}^{\mathrm{syn}} &= \int_{1}^{\infty}P(\nu,\,\gamma_{\mathrm{e}}) \frac{\diff n_{\mathrm{e}}}{\diff \gamma_{\mathrm{e}}}
    \diff \gamma_{\mathrm{e}} \\
    \label{eq:synchrotron absorption coefficient}
    \alpha_{\nu}^{\mathrm{syn}}&=-\frac{c^2}{8\pi\nu^2 m_{\mathrm{e}}c^2}\int\diff \gamma_{\mathrm{e}}
    P(\nu,\,\gamma_{\mathrm{e}})\gamma_{\mathrm{e}}^2\frac{\diff}{\diff \gamma_{\mathrm{e}}}\left[\frac{1}{\gamma_{\mathrm{e}}^2}
    \frac{\diff n_{\mathrm{e}}(\gamma_{\mathrm{e}})}{\diff \gamma_{\mathrm{e}}}
    \right].
\end{align}

For an isotropic and homogeneous medium, the synchrotron luminosity  $L_{\mathrm{\nu}}$ becomes \citep{Dermer2009}
\begin{align}
\label{eq:synchrotron luminosity}
    L_{\mathrm{\nu}} =   \frac{2\pi R_{\mathrm{sh}}^3j_{\nu}^{\mathrm{syn}}}{\tau_{\nu}} \left(
    1 - \frac{2}{\tau_{\nu}^2}\left\{
    1 - (1 + \tau_{\nu})\exp{(-\tau_{\nu})}
    \right\}
    \right).
\end{align}
Here, $\tau_{\nu} = 2\alpha_{\nu}R_{\mathrm{sh}}$ represents the synchrotron self-absorption opacity.

\subsection{Free-Free Absorption by \HII{}  Regions}
\label{subsec:free-free}
In star-forming galaxies, free-free absorption (FFA) features by \HII{} regions are typically observed around $\sim10$ MHz to $\sim10$ GHz \citep{Lacki2012}. In this paper, we consider the FFA effect both in the SAM and within the host galaxy. 

\HII{} regions, primarily composed of ionized hydrogen, induce FFA. The optical depth of FFA is given as \citep{Mezger1967}
\begin{align}
    \label{eq:FFA optical depth}
    \begin{split}
        \tau_{\mathrm{ff}} &= 
        3.014 \times 10^{-2} T_{\mathrm{e}}^{-1.5}
        \left(
        \frac{\nu}{\mathrm{GHz}}\right)^{-2.0} \left(
        \frac{EM}{\mathrm{pc\cdot cm^{-6}}}
        \right)\\
        &\quad \times\left(
        1.5 \ln{T_{\mathrm{e}}}+
        \ln{\left[4.955\times10^{-2} \left(\frac{\nu}{\mathrm{GHz}}\right)^{-1}\right]}
        \right),
    \end{split}
\end{align}
where $T_{\mathrm{e}}$ and $EM$ denote the temperature of thermal electrons and emission measure, respectively. 

From the Rankine-Hugoniot relation, the temperature of SAM, $T_{\mathrm{SAM}}$, becomes 
\begin{equation}
    \label{eq: temperature of H II region in wind-shock region}
    T_{\mathrm{SAM}} = 2 \times 10^8\,\mathrm{K} \left(
    \frac{v_{\mathrm{sh}}}{3400\,\mathrm{km\,s^{-1}}}
    \right)^2.
\end{equation} 
Using Eq.~\ref{eq: mass of swept up matters}, the emission measure of the \HII{} region in SAM is given by
\begin{equation}
    \label{eq: emission measure of wind shock region}
    \begin{split}
        EM_{\mathrm{SAM}} &= \left(\frac{M_{\mathrm{sh}}}{4\pi m_{\mathrm{p}}R_{\mathrm{sh}}^3}\right) ^{2} R_{\mathrm{sh}} \\
        &\simeq 10^4 \,\mathrm{pc\cdot cm^{-6}}\,\left(\frac{n_{\mathrm{H},\,0}}{10\,\mathrm{cm^{-6}}}\right)^2 \left( \frac{R_0}{100\,\mathrm{pc}}\right)^2 \left(\frac{R_{\mathrm{sh}}}{100\,\mathrm{pc}}\right)^{-1},
    \end{split}
\end{equation}
where we set the thickness of the \HII{} region in SAM as $R_{\mathrm{sh}}$.

For the host galaxy FFA effect, we focus on typical AGN host galaxies which exhibit star-forming characteristics \citep{Kauffmann2003}. We assume that the size of the \HII{} region in galaxies $D_{\mathrm{H II}}$ is inversely proportional to the density of ionized electrons $n_{\mathrm{H II}}$ as \citep{Hunt2009},
\begin{equation}
    \label{eq:size-density relation of H II region}
    n_{\mathrm{H II}} = 10 \mathrm{\,cm^{-3}} \left(
    \frac{D_{\mathrm{H II}}}{100\,\mathrm{pc}}
    \right)^{-1}.
\end{equation}
This leads to the emission measure of the host galaxy \HII{} regions being
\begin{equation}
    \label{eq: emission measure of host galaxy}
    EM_{\mathrm{host}} = 10^4\,\mathrm{pc\cdot cm^{-6}}\,\left( \frac{D_{\mathrm{H II}}}{100\,\mathrm{pc}}\right)^{-1}.
\end{equation}
For our calculations, we assume a typical temperature of $10^4\,\mathrm{K}$ for the host-galaxy \HII{} regions. 

When the line-of-sight \HII{} region size is smaller than the SAM size, the effective optical depth decreases by a factor of  $(D_{\mathrm{H II}}/R_{\mathrm{sh}})^2$. According to Eq.~(\ref{eq:FFA optical depth}) and (\ref{eq: emission measure of host galaxy}), a relation between the effective optical depth $\tau_{\mathrm{ff,eff}}$ and \HII{} region size are shown as 
\begin{equation}
\label{eq: dependence of FFA optical depth}
    \tau_{\mathrm{ff,eff}} = \left\{
    \begin{array}{rll}
    \left(\frac{{D_{\mathrm{H II}}}}{ {R_{\mathrm{sh}}}}\right)^2 \tau_{\mathrm{ff}} & \propto D_{\mathrm{HII}}     &  (D_{\mathrm{HII}} < R_{\mathrm{sh}}),\\
    \tau_{\mathrm{ff}} & \propto D_{\mathrm{HII}}^{-1}     &  (D_{\mathrm{HII}} \geq R_{\mathrm{sh}}).
    \end{array}
    \right. 
\end{equation}
This implies that $\tau_{\mathrm{ff,eff}}$  host \HII{} regions reaches its maximum when the size of the \HII{} region coincides with that of the SAM.

\begin{figure}
    \centering
    \includegraphics[width=\linewidth]{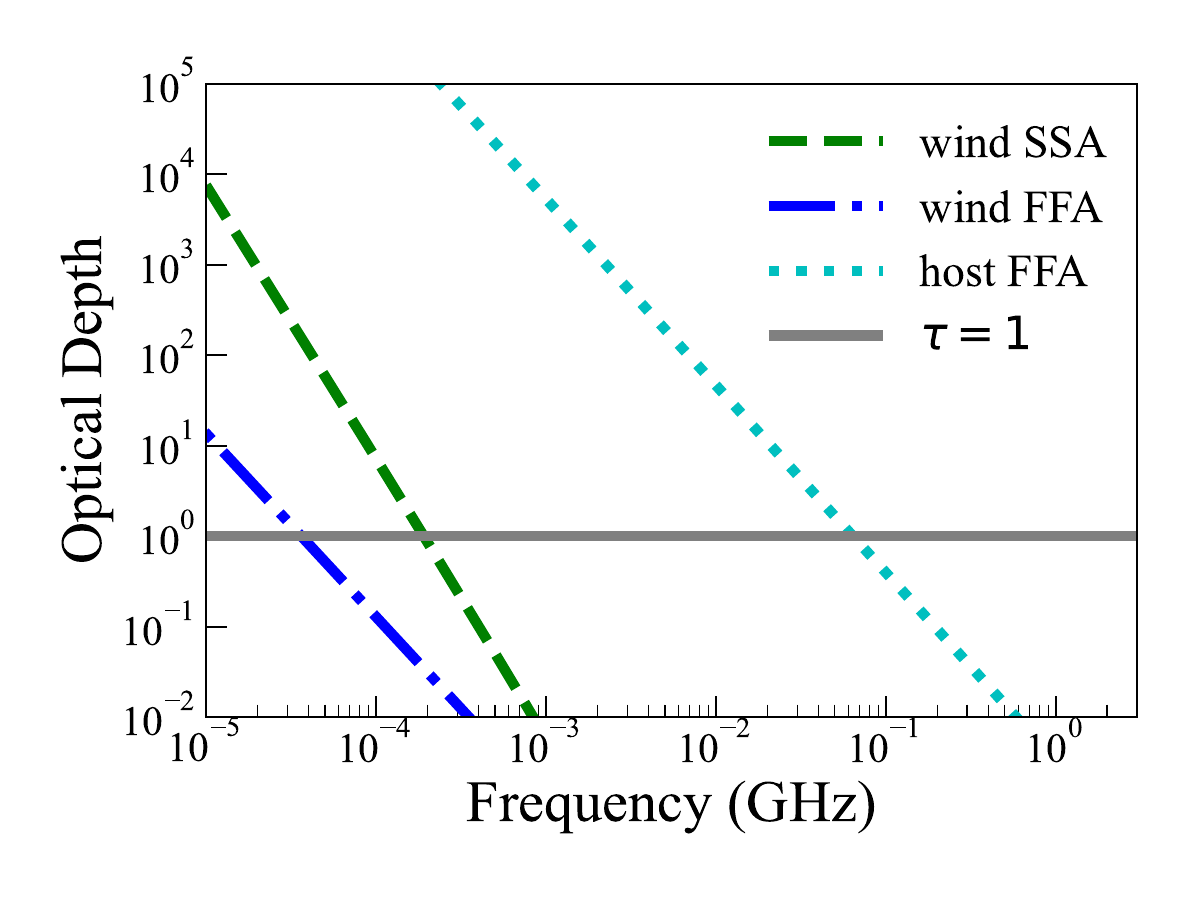}
    \caption{Optical depth of FFA by SAM (blue dash-dot) and by host-galaxy (cyan dot) comparing with that of SSA of SAM(green dash) are shown. The grey horizontal line corresponds to optical depth unity. The parameters used in these calculations are the same as Figure~\ref{fig:cooling timescale}. The sizes and temperature of the host-galaxy \HII{} regions are set to $D_{\mathrm{H II }} = 100~ \mathrm{pc}$ and $10^4\,\mathrm{K}$, respectively.}
    \label{fig:Optical depth}
\end{figure}

Figure~\ref{fig:Optical depth} shows the opacities of the combined FFA effects of the SAM and the host galaxy, in addition to the SSA effect. Adopted parameters are summarized in Table~\ref{tab:fiducial parameter}. In this figure, we set the \HII{} region to SAM size, 100 pc. The FFA of host galaxy becomes dominant, in which the FFA frequency is $\approx0.1$~GHz. Altering the size of the \HII{} region away from the SAM size leads to a decrease in optical depth, consequently causing the FFA frequency to shift towards lower frequencies. Given that our radio observational dataset for the sample is confined to frequencies above 0.1~GHz, the precise size of the \HII{} region does not exert a substantial influence on our analysis, leading to our decision to fix its size at SAM size for the purposes of this study.

\section{Result}
\label{sec:result}

\begin{figure}
    \centering
    \includegraphics[width=\linewidth]{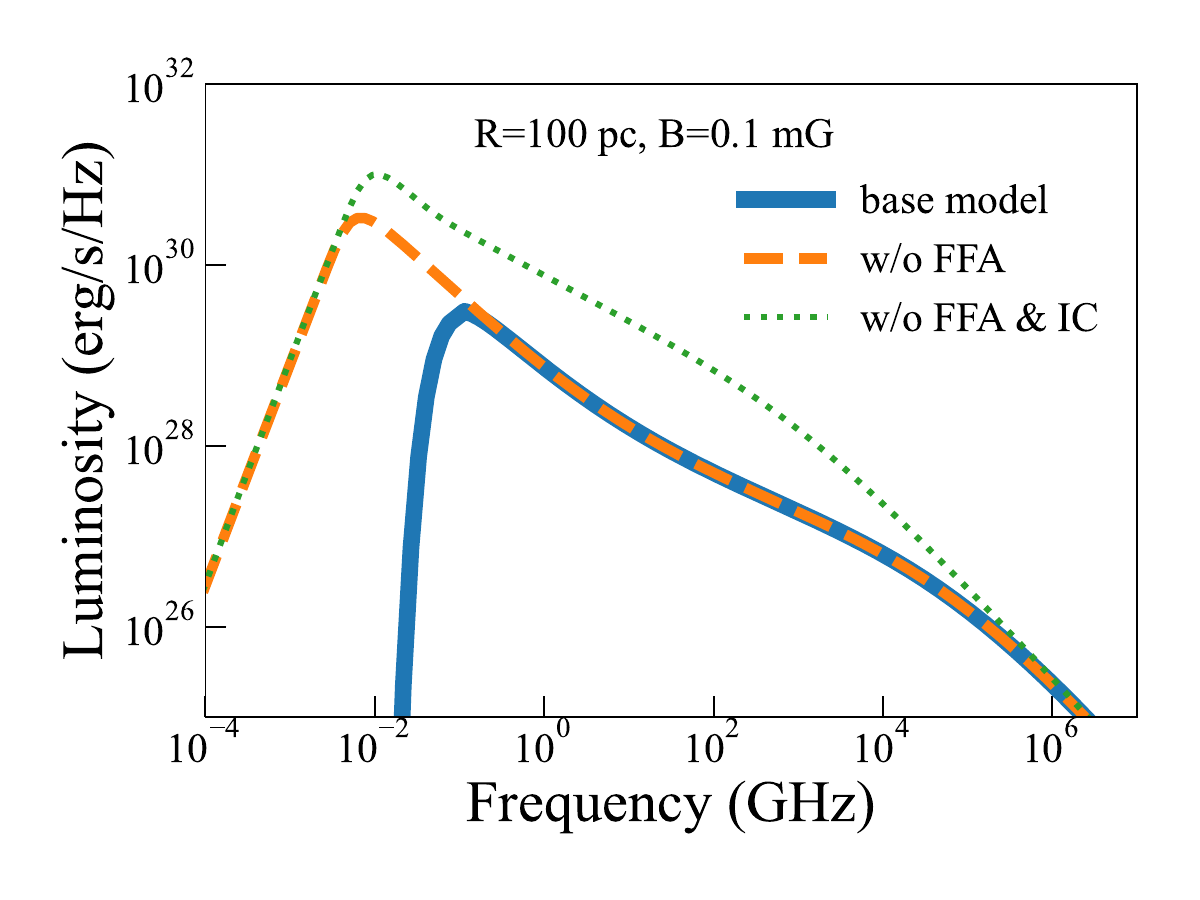}
    \caption{The synchrotron spectrum predicted by the model described in Section~\ref{sec:model} are shown. These spectra are calculated on the base model which contains host galaxy FFA and IC cooling (blue solid), the model that does not contain host galaxy FFA but does IC cooling (orange dash), and the model that does not contain either host galaxy FFA and IC cooling (green dot). We set the size of SAM region as $R_{\mathrm{sh}}=100~$pc, the magnetic field strength as $B=0.1~$mG, the bolometric luminosity of AGN as $L_{\mathrm{AGN}} = 10^{46}~\mathrm{erg \, s^{-1}}$, the velocity of UFOs as $v_{\mathrm{UFO}} = 0.1~c$, the density of ISM at $100$~pc as $n_{\mathrm{H},\,0}=10~\mathrm{cm^{-3}}$, the power law index of the spatial density distribution of the ISM as $\alpha=1$, the Bohm parameter as $\eta=1$, and the temperature of AGN accretion disk as $10$~eV. The sizes and temperature of the host-galaxy \HII{} regions are fixed to $D_{\mathrm{H II}} = 100~ \mathrm{pc}$ and $10^4~\mathrm{K}$, respectively.}
    \label{fig:synchrotron SED}
\end{figure}
\subsection{Fiducial Synchrotron Spectra}
We present three distinct synchrotron spectra from SAM regions in Figure~\ref{fig:synchrotron SED}, using the fiducial parameters outlined in Table~\ref{tab:fiducial parameter}. The three configurations include the base model, a model without FFA, and a model without both FFA and IC cooling.

For the base model, we account for both FFA from host galaxy \HII{} regions and IC cooling, setting the \HII{} region size and temperature at $D_{\mathrm{HII}}=100$~pc and $T=10^{4}$~K, respectively. Given the dominance of the host galaxy FFA (as illustrated in Figure~\ref{fig:Optical depth}), an FFA cutoff emerges around $10^{-2}~$GHz. If the \HII{} region exceeds $100$~pc, the FFA frequency will be $\nu_{\mathrm{FFA}}\leq 0.1$~GHz (see Eqs.~\ref{eq:FFA optical depth} and \ref{eq:size-density relation of H II region}). Since all observed radio data from our sample are above $0.1$ GHz, our results are not changed by fixing the \HII{} region size of host galaxies to $100$~pc.

Without considering the host galaxy FFA, an SSA cutoff appears around the MHz range. This is due to SSA being the second dominant absorption mechanism, shown in Figure~\ref{fig:Optical depth}. The SSA frequency varies depending on whether IC cooling is included, since the energy distribution of non-thermal electrons $\diff n_{\mathrm{e}}/\diff \gamma_{\mathrm{e}}$ is altered by IC cooling.

Below the THz regime, spectra generated with IC cooling are dimmer than those without. This is attributed to electrons in the Thomson regime undergoing efficient IC cooling, as demonstrated in Figure~\ref{fig:cooling timescale}.

\subsection{Comparison with Observations of RQ AGNs}
\begin{figure*}
    \centering
    \includegraphics[height=\linewidth]{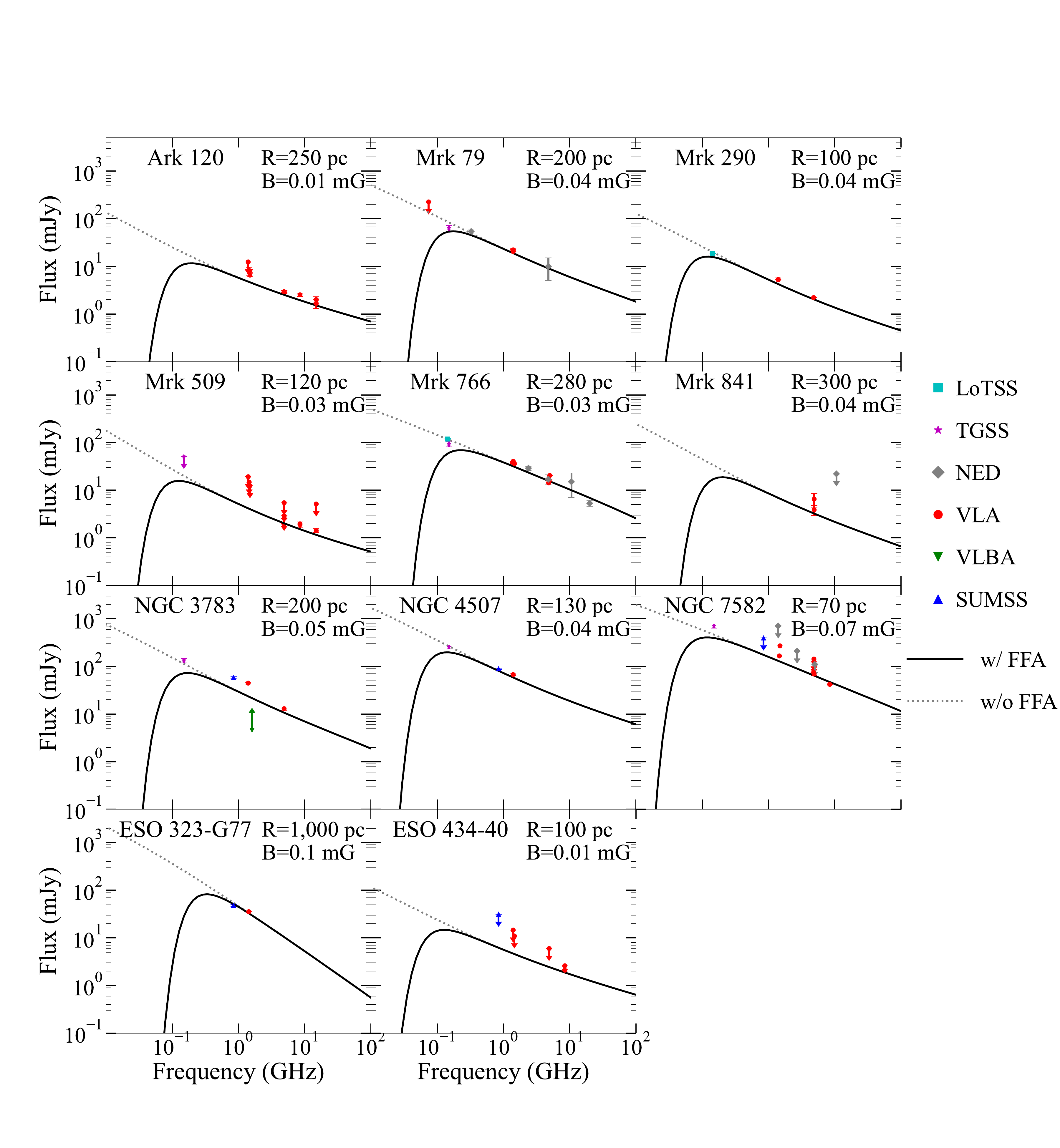}
    \caption{The cm spectrum of 11 nearby RQ AGNs. Red circles are the VLA data, green inverted triangles are the Very Long Baseline Array (VLBA) data, blue triangles are the Sydney University Molonglo Sky Survey (SUMSS) data, cyan squares are the LOFAR Two-metre Sky Survey (LoTSS) data, and the purple stars are the TIFR GMRT Sky Survey (TGSS) data, and the grey diamonds are from NED data. We treat these radio observational data as upper or lower limits. The solid line shows the predicted spectrum from our calculation, including the IC radiation and FFA of host galaxies, and the dotted line shows those not including the FFA of host galaxies. Predicted values of source size and magnetic field strength by this calculation are shown in the upper right in each panel and Table~\ref{tab:result}. In these calculations, we use AGN luminosity, $L_{\mathrm{AGN}}$, derived by X-ray observations and UFO velocity, $v_{\mathrm{UFO}}$, measured by \citet{Tombesi2010}. The other parameters are fixed to the same values as the previous computation in Figure~\ref{fig:synchrotron SED}.}
    \label{fig:fitting result}
\end{figure*}

We identified 15 RQ AGNs with both UFO signatures and radio spectral data \citep{Tombesi2010}. This radio data was taken from the NASA/IPAC Extragalactic Database (NED). Out of these, our AGN wind shock model can explain the radio spectra of eleven RQ AGNs. Figure~\ref{fig:fitting result} shows these eleven spectra, as computed by our base model, alongside their respective radio data. Detailed model parameters of these 11 sample objects, shown in Figure~\ref{fig:fitting result}, are provided in Table~\ref{tab:observational data}, with additional information in Appendix~\ref{app: observational information}. However, we note that 
the limited number of radio data points, combined with the non-uniform beam sizes across data sets, hampers is to conduct statistical model fitting. Acquiring data with uniform beam sizes across multiple frequencies in the future will refine and constrain the wind model.

We treat magnetic field strength, $B$, and source size, $R_{\mathrm{sh}}$, as variable parameters. The estimated magnetic field strength and source size are shown in each panel of the figure and summarized in Table~\ref{tab:result}. We fix the power law index of injected non-thermal electrons, the fraction of UFOs kinetic energy channeled to non-thermal electrons, and the gyrofactor to values of $p=2$, $\xi = 0.01$, and $\eta = 1$, respectively. We maintain the \HII{} region size and temperature at $D_{\mathrm{HII}}=100~$pc and $T=10^{4}$~K, same as Figure~\ref{fig:synchrotron SED}. 

We consider the observational radio data as the upper limits of AGN wind shock radiation, because the beam sizes of our radio data are larger than kpc-scale, except for VLBA data of NGC~3783 and NGC~4151. Given that the actual radio flux from SAM could be dimmer than predicted, the estimated magnetic field strength and source size in Table~\ref{tab:result} might be overestimated. The VLBA data were treated as lower limits of SAM radiation because their pc-scale beam sizes are smaller than the estimated SAM regions. These high-resolution observations have fluxes about an order of magnitude smaller than those capturing large-scale radio emission, suggesting the existence of extended radio sources than VLBA region \citep{Nagar2005,Orienti2009}. 

While realistic accelerated particle spectrum can be softer than $p=2$, as suggested by observations of supernova remnants (e.g., \citet{Aharonian2007}), the limited radio observation data prevents us from determining the value of $p$ as a variable parameter.  Further simultaneous multi-wavelength, high-resolution observations are demanded to determine the value of $p$.

The velocity of UFOs, $v_{\mathrm{UFO}}$, and the bolometric luminosity of AGNs, $L_{\mathrm{AGN}}$, are fixed to each observational result. We use the velocity of UFOs measured by \citet{Tombesi2010}. To compute $L_{\mathrm{AGN}}$, we employ the bolometric correction \citep{Marconi2003,Ichikawa2016} given by
\begin{equation}
    \label{eq:bolometric correction}
    \log L_{\mathrm{AGN}} = 0.0378(\log L_{14-195})^2 -2.03\log L_{14-195} + 61.6,
\end{equation}
where $L_{14-195}$ is the X-ray luminosity at 14--195~keV \citep{Oh2018}. These fixed parameters for respective sources are described in Table~\ref{tab:observational data}.

The predicted magnetic field strength $B$ is $\sim 0.1~\mathrm{mG}$, requiring magnetic field amplification. The magnetic field strength might be amplified by DSA in the SAM region, given the similarity in shock velocity, $v_{\mathrm{sh}}$, to that of supernova remnants. \citet{Bell2013} proposed that non-resonant instability could amplify the magnetic field to about $0.5~$mG in the shock of supernova remnants (but see also \citealp{Inoue2021a}). 

Among rest four objects, the radio spectra of NGC~3516 and NGC~4151 are shown in Figure~\ref{fig:SED of jetted object}. While, as shown in the figure, AGN wind shocks might make a significant contribution in their radio fluxes, observations suggest that their radio emission is dominated by jets, see the details in Section \ref{subsec: jet}. We could not reproduce the spectra of the remaining two objects, namely Mrk~279 and NGC~4051.

\begin{table*}
    \caption{The observational information about 11 RQ AGNs from our sample}
    \begin{tabular}{c|ccccccc}
    \hline \hline
    object & $z$ & $L_{\mathrm{X}}$ & $L_{\mathrm{AGN}}$ & $v_{\mathrm{UFO}}$ & $F_{160~\mathrm{\mathrm{\mu m}}}$ & $F_{1.4~\mathrm{GHz}}$ & Beam size~(IR)\\
     & & $\mathrm{(erg\,s^{-1})}$ & $\mathrm{(erg \, s^{-1})}$ & $(c)$& (Jy) & (mJy) & (arcsec)
    \\
    & (1) & (2) & (3) & (4) & (5) & (6) & (7)\\
    \hline
    Ark~120&0.03271&1.78e+44&6.13e+45&0.269&1.18&12.4&22\\ 
    Mrk~79&0.02221&4.79e+43&1.12e+45&0.091&2.37&22.0&30\\ 
    Mrk~290&0.03023&4.79e+43&1.12e+45&0.141&0.116&5.32&22\\ 
    Mrk~509&0.03440&2.75e+44&1.09e+46&0.172&1.23&19.2&22\\ 
    Mrk~766&0.01288&9.77e+42&1.55e+44&0.091&2.94&38.1&22\\ 
    Mrk~841&0.03642&1.02e+44&2.98e+45&0.034&0.182&$<$14.8&  22\\
    NGC~3516&0.00884&1.95e+43&3.63e+44&0.001&1.20&31.0&22\\ 
    NGC~3783&0.00973&3.63e+43&7.90e+44&0.013&4.49&44.6&40\\ 
    NGC~4151&0.00333&1.48e+43&2.58e+44&0.105&9.06&348&78\\ 
    NGC~4507&0.01180&5.75e+43&1.42e+45&0.177&4.68&67.4&35\\ 
    NGC~7582&0.00525&4.90e+42&6.72e+43&0.255&77.3&270&87\\ 
    ESO~323-G77&0.01501&1.55e+43&2.73e+44&0.005&7.53&35.5&  22\\
    ESO~434-40&0.00849&3.39e+43&7.24e+44&0.118&0.477&14.6&22\\ 
    \hline
    \end{tabular}
    \\ 
    \raggedright
    Col. (1) redshift which is extracted from NASA/IPAC Extragalactic Database (NED); Col. (2) $14-195~\mathrm{keV}$ X-ray luminosity \citep{Oh2018}; Col. (3) bolometric luminosity calculated by Eq (\ref{eq:bolometric correction}); Col. (4) UFOs velocity determined by the observations of blueshifted Fe absorption line \citep{Tombesi2010}; Col. (5) IR flux at $160~\mu$m based on NED in the unit of Jy; Col. (6) radio flux from NVSS data in the unit of mJy; Col. (7) Beam size of IR observation in the unit of arcsec.
    \label{tab:observational data}
\end{table*}

\begin{table}
    \centering
    \caption{The predicted parameters of our sample.}
    \begin{tabular}{c|cccc}
    \hline \hline
    object name & $R_{\mathrm{sh}}$ & $\theta$ & $B$ & $\mathcal{R}_{\mathrm{SF}}$ \\
     & (pc) & (arcsec) & (mG) &$-$
    \\
    & (1) & (2) & (3) & (4) \\
    \hline
    Ark~120&250&0.3791&0.01&0.38\\
    Mrk~79&200&0.4411&0.04&0.44\\
    Mrk~290&100&0.1636&0.04&0.09\\
    Mrk~509&120&0.1734&0.03&0.26\\
    Mrk~766&280&1.0531&0.03&0.31\\
    Mrk~841&300&0.4104&0.04&$>$0.05  \\
    NGC~3516&220&1.1997&0.08&0.16\\
    NGC~3783&200&0.9920&0.05&0.41\\
    NGC~4151&130&1.8695&0.03&0.11\\
    NGC~4507&130&0.5330&0.04&0.28\\
    NGC~7582&70&0.6400&0.07&1.16\\
    ESO~323-G77&1000&3.2356&0.10&0.86  \\
    ESO~434-40&100&0.5676&0.01&0.13\\
    \hline
    \end{tabular}
    \\ 
    \raggedright
    Col. (1) radio source size in the unit of pc; Col. (2) angular size in the unit of arcsec; Col. (3) magnetic field strength in the unit of mG; Col. (4) ratio of radio flux from star formation to that of $1.4~$GHz VLA observation. 
    \label{tab:result}
\end{table}

\section{Discussion}
\label{sec:discussion}
\subsection{Alternative Sources of Radio Emission}

The origins of radio emissions in RQ AGNs are still a mystery. Possible candidates other than AGN disk wind are star formation, AGN corona, and weak jets (see \citealp{Panessa2019} for review). We discuss these possibilities in this section.

\subsubsection{Star Formation Activity}
Star formation can contribute to the radio flux in RQ AGNs. The radio-IR correlation observed in star-forming galaxies suggests that a portion of the radio emission in RQ AGNs may arise from star formation \citep{Wunderlich1987, Sudoh2021PhRvD.103h3017S}. A similar radio-IR correlation is observationally seen in RQ AGNs \citep{Sopp1991,Sargent2010}. Using the radio-IR correlation, we can estimate the radio luminosity from star formation \citep{Bell2003}:
\begin{equation}
    \label{eq: estimated star-forming radio flux from IR flux}
        \log L_{\mathrm{radio,\,SF }}=26.4687 
        +1.1054 \log \left(\frac{L_{\mathrm{IR},\,\mathrm{SF}}}{\mathrm{L}_{\odot}}\right).
\end{equation}
Here, $L_{\mathrm{radio,\,SF }}$ and $L_{\mathrm{IR,\,SF}}$ are the monochromatic radio luminosity at $1.4$~GHz and the total IR luminosity integrated between $8~\mu$m and $1000~\mu$m, respectively. We derive the total IR luminosity following  \citet{Zakamska2016}, using the IR flux at $160~\mu$m observed by the Photodetector Array Camera and Spectrometer (PACS) on the Herschel Space Observatory \citep{Poglitsch2010A&A...518L...2P}.

Using this correlation, we can compute the ratio of the radio flux from star-formation activity to observed radio flux using Eq.~\ref{eq: estimated star-forming radio flux from IR flux} as 
\begin{equation}
    \mathcal{R}_{\mathrm{SF}} \equiv L_{\mathrm{radio,SF }} / L_{1.4\mathrm{GHz}}.
\end{equation}
We use the VLA data at $1.4$ GHz based on NED for $L_{1.4\mathrm{GHz}}$. The observational data and the ratio $\mathcal{R}_{\mathrm{SF}}$ for our sources are summarized in Table~\ref{tab:observational data} and \ref{tab:result}, respectively. 

 Our analysis indicates that star-forming radio fluxes of all objects except NGC~7582 are insufficient to explain their observed radio flux, i.e., $\mathcal{R}_{\mathrm{SF}} < 1$. NGC~7582 hosts diffuse radio sources extending to $\sim$~kpc exist \citep{Orienti2009} coexisting together with near-IR counterparts \citep{Mezcua2015}. Extended star-formation activity would significantly contribute to the radio flux in the case of NGC~7582, although we can still model their radio spectrum with our AGN wind shock model (see Figure~\ref{fig:fitting result}). As discussed in Section~\ref{subsec:ngVLA}, future spatially high-resolution radio observations will be able to distinguish these two scenarios for NGC~7582.

\subsubsection{AGN Corona Activity}
There is evidence suggesting a possible relation between X-ray and millimeter radio luminosities in RQ AGNs \citep{Laor2008MNRAS.390..847L, Behar2015MNRAS.451..517B, Doi2016, Inoue2018a, Inoue2020ApJ...891L..33I, Kawamuro2022, Baldi2022MNRAS.510.1043B, Ricci2023arXiv230604679R, Michiyama2023arXiv230615950M}. Given that RQ AGN X-ray emissions arise from Comptonization in hot coronae above accretion disks \citep{Haardt1991,Liu2003, Jin2012MNRAS.420.1825J, Morgan2012ApJ...756...52M}, this radio-X relation suggests that radio emission is also generated in the corona scale, likely due to coronal synchrotron emission \citep{DiMatteo1997, Inoue2014, Raginski2016}. However, recent high spatial-resolution radio observations indicate that coronal radio emission can not extend down to the cm band due to the coronal SSA effect at around a few tens GHz \citep{Inoue2018a, Inoue2020ApJ...891L..33I, Kawamuro2022, Michiyama2023arXiv230615950M, Michiyama2024arXiv240400647M}. Thus, for the frequency range considered in this study ($\lesssim10$~GHz), the contribution from coronal synchrotron emissions is minimal.

\subsubsection{Weak Jet Activity}
\label{subsec: jet}
Some of the nearby RQ AGNs exhibit narrow radio jet features, extending to several tens of pc to kpc, whose scales are smaller than those of RL AGNs \citep{Ho2001, Nagar2001, Michiyama2022}. While most of our RQ AGN samples show unresolved and compact radio morphologies, i.e. no clear evidence for jets, NGC~3516 and NGC~4151 display jet-like elongated features. 

NGC~3516 has linear-shape outflow extending to several kpc \citep{Miyaji1992,Baum1993}, where the radio morphology of the central several hundred pc consists of a flux-dominated central core of $<18$~pc and a number of linearly extending blobs. 

NGC~4151 shows a two-sided jet structure spanning around 300~pc with six bright knots confirmed by 5 GHz and 8 GHz observation by VLA \citep{Pedlar1993} and $1.51$ GHz observation by enhanced Multi-Element Radio Linked Interferometer Network (eMERLIN) \citep{Williams2017}.

\begin{figure*}
    \centering
    \includegraphics[width=\linewidth]{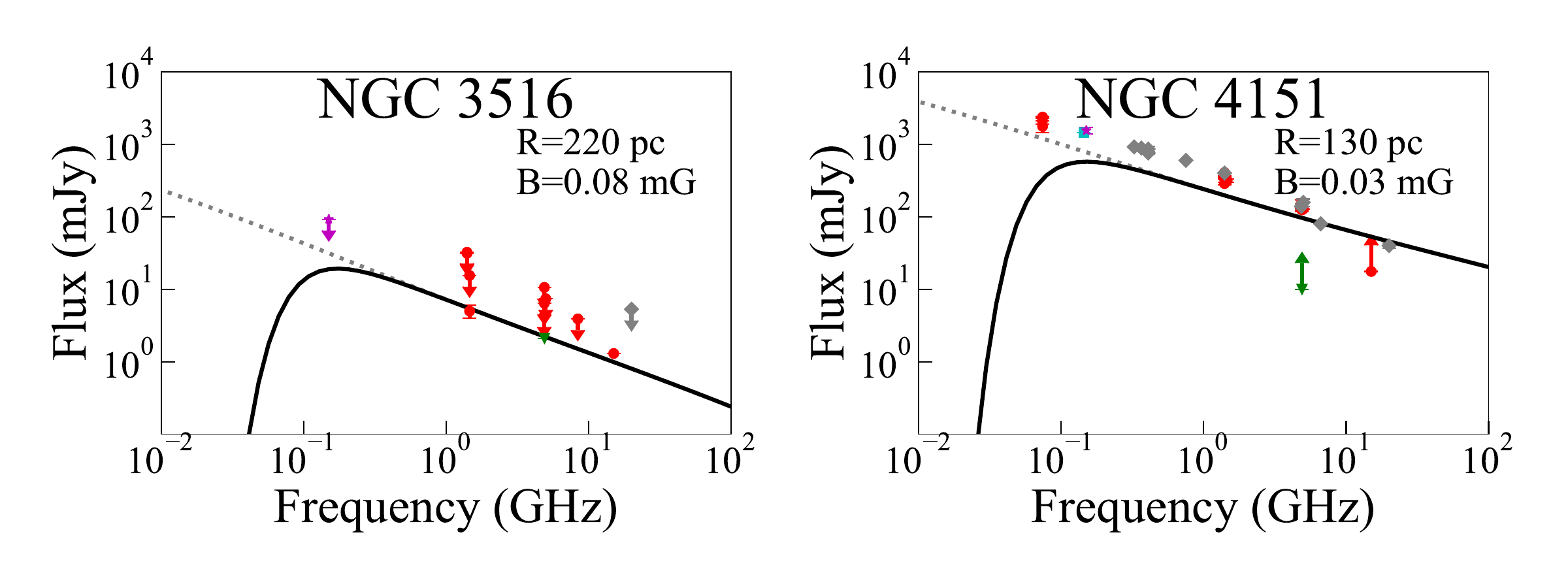}
    \caption{SED of two jetted objects, NGC~3516 and NGC~4151. Radio observational data, predicted flux, and the parameters used in the calculations are shown in the same way as Figure~\ref{fig:fitting result}. Though the spectra of these two objects are modeled by our AGN wind shock model, the predicted sizes are not consistent with the observational radio morphology described in Section \ref{subsec: jet}.}
    \label{fig:SED of jetted object}
\end{figure*}

Figure~\ref{fig:SED of jetted object} shows the observed flux of NGC~3516 and NGC~4151 compared to our AGN wind shock model. The estimated source sizes are $100$~pc, but observationally there are no such wind-like structures other than jet structures. Although our AGN wind shock model can model the spectra of these galaxies, their radio emissions likely stem from jet activities.

\subsection{Future Radio Observational Tests for the AGN wind Shock Scenario}
\label{subsec:ngVLA}
In this study, we introduce a AGN wind shock model to elucidate the radio spectra of sample RQ AGNs. Here, we discuss potential strategies to validate this hypothesis, leveraging high spatial resolution and time variability observations.

Our model predicts radio-emitting regions of approximately $100$~pc, corresponding to 0.1--1~arcseconds with a radio flux of 1--100~mJy at 1--10~GHz for our nearby UFO samples. As the other possible scenarios (e.g., star formation, corona, and jet) would predict different morphologies, the spatially resolved structure would be a crucial test.

While it would be challenging to resolve the shocked regions at 1.4~GHz by VLA \citep{Condon1998a}, higher frequencies, such as 30GHz, appear more promising according to the VLA sensitivity calculator\footnote{\url{https://obs.vla.nrao.edu/ect}}. Moreover, the next-generation Very Large Array (ngVLA, \citet{Murphy2018}), with its enhanced capabilities, would resolve these regions even at 1.4~GHz, as indicated by the ngVLA sensitivity calculator \footnote{\url{https://gitlab.nrao.edu/vrosero/ngvla-sensitivity-calculator}}. 

Radio time variability observations can also provide insights into the origin of the radio source. The scale of observed variability can provide estimates of the size of unresolved radio sources. Furthermore, by examining the correlation with time variability in other wavelength ranges, we would be able to identify the radio source indirectly. 

From our spherically symmetric one-zone model the radio-emitting regions are predicted as $\sim100$~pc scale, corresponding time variability of several hundreds of years scale. However, observations have already shown the evidence of AGN winds with time variations of $\sim10$~days \citep{Reeves2020} to $\sim$year \citep{Ogorzalek2022a}. The exact nature of RQ AGN time variability remains elusive \citep{Neff1983,Wrobel2000,Mundell2009,Jones2011,Baldi2015a,Chen2022,Panessa2022}. Further radio observations of time variability are necessary to identify the origin of the radio source through time variation.

\subsection{Cosmic-Ray Protons Acceleration}
In addition to non-thermal electrons, high-energy cosmic-ray protons are accelerated via the DSA mechanism (e.g., \cite{Abe2022}). Given the significantly larger mass of protons compared to electrons, their synchrotron and IC cooling are less efficient. The maximum energy of cosmic-ray protons is estimated by the relation $t_{\mathrm{acc}}=t_{\mathrm{dyn}}$. Using the Eq.~\ref{eq:dynamical timescale} and \ref{eq:acceleration timescale} the maximum energy of the protons $E_{\mathrm{p,\,max}}$ can be derived as 
\begin{align}
    E_{\mathrm{p,\,max}} &= \frac{3 e B v_{\mathrm{sh}} R_\mathrm{sh}}{8 \eta c} \\
    &\simeq 100 \mathrm{~PeV} \left(\frac{B}{0.1 \mathrm{~mG}}\right)\left(\frac{v_\mathrm{sh}}{3400 \mathrm{~km} / \mathrm{s}}\right)\left(\frac{R_\mathrm{{sh}}}{100 \mathrm{~pc}}\right).
\end{align}

These cosmic-ray protons can produce secondary electrons through hadronic interactions. However, these secondary leptons would not significantly influence the radio spectrum. The energy flux of secondary electrons, denoted as $P_{\mathrm{e,\,2nd}}$, can be explicated as follows:
\begin{equation}
    P_{\mathrm{e,\,2nd}} = R_{\mathrm{e,\,pp}}  f_{\mathrm{pp}}  P_{\mathrm{p,\,1st}} \simeq 1.5 \times 10^{-4} L_{\mathrm{kin}},
\end{equation}
where $R_{\mathrm{e,\,pp}} \simeq 0.1$ embodies the energy ratio of secondary electrons engendered through pp interactions relative to the total energy emitted by pp interactions. The parameter $f_{\mathrm{pp}}$ is expressed as $f_{\mathrm{pp}} = t_{\mathrm{dyn}}/t_{\mathrm{pp}} \simeq 1.5 \times 10^{-2}$, wherein $t_{\mathrm{pp}}$ signifies the cooling timescale of pp interactions with $4n_\mathrm{H}(R_\mathrm{sh})$ under strong shock assumption \citep{Kelner2006}. $P_{\mathrm{p,\,1st}}$ designates the injected energy of non-thermal protons. Here, we postulate that 10\% of the AGN kinetic luminosity transforms into the energy of non-thermal protons. As the injected energy of primary electrons is $P_{\mathrm{e,\,1st}} = 10^{-2} L_{\mathrm{kin}}$, the contribution of secondary electrons can be disregarded.

\section{Conclusion}
\label{sec: conclusion}
In this study, we developed a comprehensive model to understand the radio emissions observed in RQ AGNs. Our approach was grounded in the framework of a spherical, one-zone, self-similar expansion model of energy-conserving outflow, as proposed by \citet{Nims2015}.

Our analytical model posits that UFOs, originating from the central SMBH, collide with the ISM. This collision leads to the acceleration of non-thermal electrons via the DSA mechanism at the SAM region. By solving the transport equation and accounting for synchrotron and IC cooling, we derived the energy density spectrum of these electrons. This, in turn, allowed us to compute the resultant radio synchrotron spectrum.

We compare the calculated radio spectrum with data from 15 RQ AGNs known to possess UFOs. To calculate the model spectrum, we use bolometric luminosity of AGN derived by X-ray observation \citep{Oh2018}, and the velocity of UFOs measured by \citet{Tombesi2010}. By treating the source size and magnetic field strength as adjustable parameters, we were able to determine the values that model the observed radio spectral data. 11 AGNs among the whole sample would be explained by the AGN wind shock model, with typical source sizes around $R_{\mathrm{sh}}\sim 100$~pc. The inferred magnetic field strength, $B\sim0.1$~mG, and shock front velocity, $v_{\mathrm{sh}}\sim 3400~$km/s, were consistent with those observed in supernova remnants.

Our analysis, leveraging the radio-IR correlation and IR data from PACS, indicated that star formation alone cannot account for the radio flux in most objects, with the notable exception of NGC~7582. For NGC~7582, the radio flux, as predicted by the radio-IR correlation, aligns closely with VLA observations. Given the presence of diffuse radio sources extending to about kpc \citep{Orienti2009} and the identification of a near-IR counterpart \citep{Mezcua2015}, we cannot dismiss the contribution of star formation to its radio flux. 

While AGN disk wind component make a significant contribution to the radio spectra of NGC~3516 and NGC~4151, these galaxies primarily exhibit jet structures. The absence of confirmed AGN wind shock structures yet in these galaxies suggests that our AGN wind shock model might not fully account for their radio emissions.

The predicted source regions for all sample objects are within the resolving capabilities of the next-generation Very Large Array (ngVLA). This presents an exciting opportunity for further radio observations to validate or refine our AGN wind shock model.

In conclusion, our study offers a robust framework to understand the radio emissions in RQ AGNs. While our AGN wind shock model aligns well with several observed spectra, there remain galaxies where other mechanisms, such as star formation or jet structures, might play a dominant role. Future high-resolution observations will be pivotal in refining our understanding and validating our proposed model.

\section*{Acknowledgements}
We thank the anonymous referee for his/her helpful comments which improved the manuscript. We also would like to thank Chris Done and Misaki Mizumoto for useful discussions and comments. This research has made use of the NASA/IPAC Extragalactic Database, which is funded by the National Aeronautics and Space Administration and operated by the California Institute of Technology. T.M. and Y.I. appreciate support from NAOJ ALMA Scientific Research Grant Number 2021-17A. T.M. is supported by JSPS KAKENHI grant No. 22K14073. Y.I. is supported by JSPS KAKENHI Grant Number JP18H05458, JP19K14772, and JP22K18277. This work was supported by World Premier International Research Center Initiative (WPI), MEXT, Japan.  

\appendix
\section{General Properties of Our UFO Sample}
\label{app: observational information}
\begin{figure*}
    \centering
    \includegraphics[width=\linewidth]{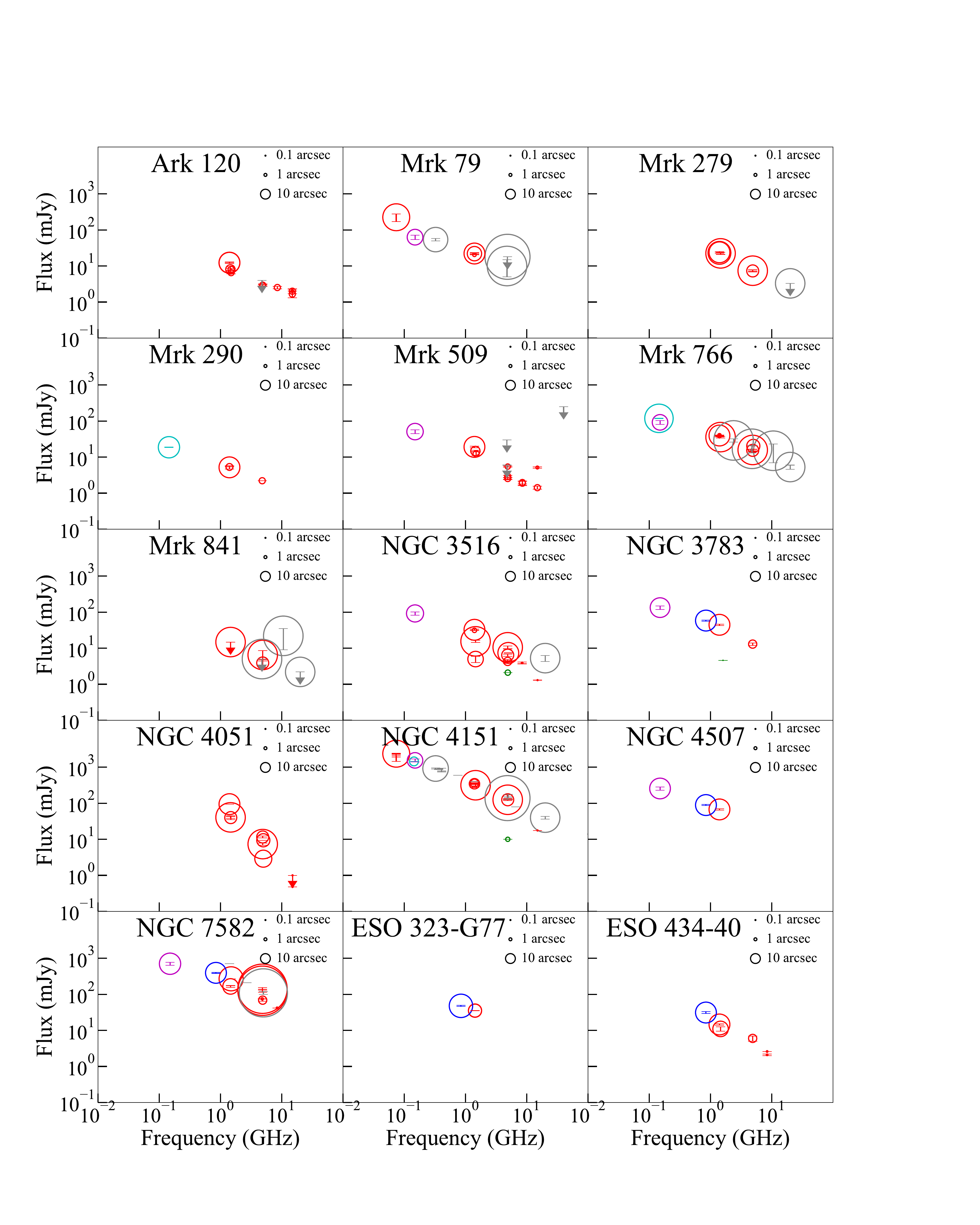}
    \caption{Radio observational data of 15 nearby RQ AGNs. Observational data are shown in the same way as Figure \ref{fig:fitting result}, but there is the beam size of each observation as opened circles.}
    \label{fig: 15 observational plot }
\end{figure*}

In this appendix, we provide in-depth supplementary information on our sample objects. The observational data for the 15 RQ AGNs is displayed in Figure \ref{fig: 15 observational plot }. Owing to the diverse scales of the observational data, in some instances, we have treated these datasets as indicative upper or lower limits for the purposes of our model predictions. 

\textbf{Ark~120:} A proximate Seyfert 1 galaxy, Ark~120 has a supermassive black hole (SMBH) with a mass of $M_{\mathrm{BH}}=(1.5 \pm 0.19) \times 10^{8} ~\mathrm{M_{\odot}}$ \citep{Peterson2004}. Notably, its X-ray spectrum lacks intrinsic absorption features, such as dust reddening in the IR band or a warm absorber in the UV and X-ray bands. As a result, Ark 120 is often referred to as a "bare AGN" \citep{Ward1987,Crenshaw1998,Patrick2011}.

\textbf{Mrk~79:} This nearby Seyfert 1 galaxy has an SMBH mass of $M_{\mathrm{BH}}=(5.2 \pm 1.4) \times 10^{7} ~\mathrm{M_{\odot}}$ \citep{Peterson2004}. In addition to the UFO features, a bipolar outflow is evident from its [O III] line \citep{Freitas2018}. Previous VLA mappings at $6$~cm and $20$~cm revealed an elongated radio structure extending up to $\sim 2$~kpc \citep{Ulvestad1984}. The radio flux from this linear structure is approximately $10.6$~mJy at $20$~cm, which is roughly half of the NRAO VLA Sky Survey (NVSS) data \citep{Condon1998a}.

\textbf{Mrk~290:} A Seyfert 1.5 galaxy, Mrk~290  has an SMBH with a mass of $M_{\mathrm{BH}}=(9.4 \pm 6.0) \times 10^{7} ~\mathrm{M_{\odot}}$ \citep{Campitiello2020}. VLA-A configuration observations at $5$~GHz found the object as a point source with a beam size of $\sim 0.5$ arcsec \citep{Kellermann1994}.

\textbf{Mrk~509:} Another Seyfert 1.5 galaxy, Mrk~509 has an SMBH mass of $M_{\mathrm{BH}}=(2.9 \pm 1.8) \times 10^{8} ~\mathrm{M_{\odot}}$ \citep{Campitiello2020}. VLA mappings at $6$~cm and $20$~cm revealed an elongated radio source extending up to $\sim 2$~kpc with a diffuse structure \citep{Singh1992}. The flux from this extended source is significantly less than the NVSS data \citep{Condon1998a}.

\textbf{Mrk~766:} This Seyfert 1 galaxy has an SMBH with a mass of $M_{\mathrm{BH}}= (1.76 \pm 1.56)\times 10^{6} ~\mathrm{M_{\odot}}$ \citep{Bentz2009}. Multiple observations, including those by VLA and GMRT, have failed to resolve its radio source. The most detailed observation by \citet{Kukula1995} determined the maximum size of the nuclear radio source to be $0.26$~arcsec, corresponding to the linear size $65$~pc.

\textbf{Mrk~841}: A Seyfert 1 galaxy, its SMBH has a mass of $M_{\mathrm{BH}}=(4.7 \pm 1.6) \times 10^{7} ~\mathrm{M_{\odot}}$ \citep{Brotherton2020}. Variable iron line, due to multiple warm absorbers, were observed from this source \citep{Longinotti2010}. VLA observations has been conducted  at a spatial resolution of up to 15~arcsec at a wavelength of 4.89 GHz \citep{Edelson1987}.

\textbf{NGC~3516:} A Seyfert 1.5 galaxy, its SMBH has a mass of $M_{\mathrm{BH}}=(4.3 \pm 1.1) \times 10^{7} ~\mathrm{M_{\odot}}$ \citep{DeRosa2018}. Observations have revealed an elongated radio morphology with several knot components~\citep{Ho2001,Miyaji1992}. The dominant radio component, named region A, has a flux of approximately $2.80$~mJy with a size of $< 0.1$~arcsec. From optical emission lines observation, \citet{Miyaji1992} found that there is an ionized region at a distance of $\sim 20$~kpc from the central nucleus, which is elongated to the central nucleus. This feature is explained by collimated ionizing photons from the nuclear relativistic jet.

\textbf{NGC~3783:} This Seyfert 1 galaxy has an SMBH mass of $M_{\mathrm{BH}}= 2.34 \pm0.43\times 10^{7} ~\mathrm{M_{\odot}}$ \citep{Bentz2021}. Observations at various frequencies have not resolved its radio source \citep{Schmitt2001, Orienti2009}. \citet{Amorim2021} found an ionized region extending to $\sim 100$~pc by colonal line observation. 

\textbf{NGC~4151:} A Seyfert 1 galaxy, its SMBH has a mass of $M_{\mathrm{BH}}=(2.1 \pm 0.55) \times 10^{7} ~\mathrm{M_{\odot}}$ \citep{DeRosa2018}. Observations have revealed a clear jet structure extending up to $\sim 300$~pc \citep{Williams2017}. The total radio flux from the jet region is $\sim 100$~mJy, which is half of that of NVSS data \citep{Condon1998a}, and this total flux is dominated by that from central radio knot, which is $72.00 \pm 0.02$~mJy.

\textbf{NGC~4507:} This object is a nearby Seyfert 2 type galaxy. This object was slightly resolved at $3.5$ cm by Australia Telescope Compact Array (ATCA) \citep{Morganti1999}.

\textbf{NGC~7582:} This Seyfert 2 galaxy has been mapped by both VLA \citep{Ulvestad1984c,Orienti2009} and ATCA \citep{Morganti1999}. Its radio morphology includes a compact nuclear component, with a size of $\lesssim40$~pc, and three extended clumps, which are believed to represent star-forming regions \citep{Orienti2009, Mezcua2015}. \citet{Morris1985} found that there is [O III] line outflow which coexists with the radio source region and the average velocity of that is $\sim 100$~km/s.

\textbf{ESO~323-G77}: This object is known to be a changing-look AGN. Accoding to VLA observations, Its radio morphology were classified into linear shape extending 1.69~kpc, but reality is not completely certain due to noisiness \citep{Nagar1999}.

\textbf{ESO~434-40 (MCG~5-23-16):} This object is a Seyfert 2 galaxy, it has been mapped by VLA and shows a slightly resolved morphology in two components \citep{Ulvestad1984c,Orienti2009}. The largest linear scale is $\sim200$~pc for $8.4$~GHz VLA observation \citep{Orienti2009}.

\bibliographystyle{aasjournal}



\end{sloppypar}
\end{document}